\documentclass[reviewcopy]{elsart}

\usepackage{epsfig}
\usepackage{amssymb}
\usepackage{amsmath}

\begin{document}
\begin{frontmatter}

\title{Field-Induced Quantum Criticality of Systems with
Ferromagnetically Coupled Structural Spin Units}

\author[SA]{I. Rabuffo\corauthref{ir}},
\author[SA]{M.~T. Mercaldo},
\author[SA]{L. De Cesare},
\author[NA]{A. Caramico D'Auria},

\corauth[ir]{Corresponding author: Tel.: +39 089 965 392; Fax: +39 089 965 275; \\
{\em E-mail address:} rabuffo@sa.infn.it}

\address[SA]{Dipartimento di Fisica ``E. R. Caianiello'',
Universit\`a di Salerno and CNISM, Unit\'a di Salerno,  I-84081 Baronissi
(Salerno), Italy}
\address[NA]{ Dipartimento di Scienze Fisiche, Universit\`a di Napoli Federico II
and ``Coherentia'' CNR-INFM, I-80125  Napoli, Italy}

\begin{abstract}
The field-induced quantum
criticality  of compounds with ferromagnetically coupled
structural spin units (as dimers
 and ladders) is explored by applying Wilson's renormalization group
framework to an appropriate effective action. We determine the
low-temperature phase boundary  and the behavior of relevant
quantities decreasing the temperature with the applied magnetic
field fixed at its quantum critical point value. In this context,
a plausible interpretation of some recent experimental results is also
suggested.
\end{abstract}

\end{frontmatter}

The study of quantum phase transitions (QPT's) and low-temperature
properties close to a quantum critical point (QCP) of a wide
variety  of materials has recently attracted considerable
attention and constitutes today a topical subject in condensed
matter physics \cite{sach99}.

The main methods of tuning a system toward a QCP are essentially
based on manipulation of doping, pressure and magnetic field.  A
lot of  experiments on spin compounds has shown that the magnetic
field is the most convenient non-thermal parameter to control the
distance from a QCP.

Emerging magnetic field-induced QPT's have been observed in
 quantum antiferromagnetic (AFM) compounds as $KCuCl_3$
\cite{shira97}, $Tl Cu Cl_3$ \cite{shira97,osa99}, $Ba Cu Si_2
O_6$ \cite{jai04}, and $Cu_2 (C_5 H_{12} N_2)_2 Cl_4$
\cite{chab97} which consist of weakly coupled low-dimensional
structural spin units such as dimers \cite{shira97,osa99,jai04} or
ladders \cite{chab97,dag99}. In absence of a magnetic field, these
systems exhibit a gap between a singlet ground state and the
lowest triplet excitation. When the field is switched on and
increased, the gap decreases by Zeeman effect  and a field-induced
QPT occurs at a critical field which measures the amplitude of the
original gap.

Recently,  also the case of weakly ferromagnetically coupled spin
units, although less explored then the AFM one, has attracted a lot of interest
\cite{gre15,mil98,dal00,zho04,uhr04,reg09} motivated by the
possible existence of dimer and ladder materials with effective
ferromagnetic (FM) interactions between the basic spin units
\cite{gre15,dal00,reg09,kat98,val02}. Inelastic neutron scattering
(INS) investigations on the spin dimer compounds $Cs_3 Cr_2 Br_9$
\cite{gre15}, on the layered cuprate $Sr_{14} Cu_{24} O_{41}$
\cite{reg09} and on $ Ca V_2 O_5$ \cite{uhr04}, which contains
layers of coupled two-leg ladders, appear particularly meaningful
in this direction. An intriguing feature is that these materials
are characterized by frustrated inter-units microscopic couplings.
Nevertheless, the INS predictions suggest that their essential
physics can be captured by means of a spin model where the
original frustrated inter-units couplings are replaced by
FM ones. A zero-temperature numerical study of this
effective simplified spin model has been performed in Ref.
\cite{dal00} and the results are in agreement with previous
conventional investigations \cite{nor98}.

Effective inter-units FM couplings may also arise from different
mechanisms. An important example has been considered in Ref.
\cite{uhr04} where a unified picture of recent INS susceptibility
data \cite{sid03} for stripe-ordered $La_{15/8} Ba_{1/8} CuO_4$ is
given based on a model of two-legs spin ladders where an effective
FM inter-ladder coupling results from integrating out the degrees
of freedom in the stripes.

Support to the  ferromagnetically coupled structural spin units
scenario is given also by first-principles and Monte Carlo
calculations of electronic structures about the nature of the
spin-singlet ground state in the dimer compound $Ca Cu Ge_2 O_6$
\cite{val02}. The results obtained for the susceptibility and magnetization
behavior give indeed  evidence of effective FM inter-dimer
couplings and of long-range interactions effects. In
addition there are indications  \cite{dal00}
of compounds which consist of
ferromagnetically coupled  ladders, as  $Sr Cu_2 O_3$ \cite{azu99}.

To gain insight into the properties of spin materials with weak FM
inter-units couplings, also in the presence of a magnetic field,
 the original complex spin Hamiltonian  is conveniently mapped
in an effective spin-1/2 FM XXZ model with
spins, exchange couplings and longitudinal magnetic field expressed
perturbatively as linear combinations of the original microscopic ones
\cite{tac70,gre15,mil98,dal00,zho04}.
The final Hamiltonian has the general structure
\begin{equation}
\label{eq1}
H =  - \sum\limits_{i,j = 1}^N {\left[ {\mathop
J\nolimits_{ij} (S_i^x S_j^x  + S_i^y S_j^y ) + K_{ij} S_i^z S_j^z
} \right]}  - h\sum\limits_{i = 1}^N {S_i^z }
\end{equation}
where $S_i^{\alpha} (\alpha = x, y, z)$ denote the effective spin
components at site $i$ (which is a spin unit index in the original
spin model) of a lattice with N sites and $J_{ij} > 0$, $K_{ij} >
0$ (with $J_{ii} = K_{ii} = 0$).

 A systematic study of the thermodynamics of weakly
ferromagnetically coupled spin units was performed \cite{tac70}
more than three decades ago within a mean field approximation
(MFA). Further few studies for specific problems at zero
temperature and, in some cases, in absence of a magnetic field,
have been achieved at numerical, Hartree-Fock and
MFA levels \cite{gre15,mil98,dal00,zho04,val02,tac70}.
However, reliable experimental and theoretical
studies of the magnetic properties and of the  phase boundary of these systems
close to the field-induced QCP are still lacking at the present time.

The aim of this letter is to give a contribution in this direction by
using an appropriate functional representation of the FM spin
model (1) \cite{dac80,car05} and a wilsonian renormalization group
(RG) approach, which is the most valid and reliable tool to
take properly into account fluctuations effects close to quantum
and classical critical points. For generality purposes  and with
the intent to interpret some experimental findings
\cite{tsu69,noh04}, we refer to a $d$-dimensional spin lattice and
include the possibility of long-range FM couplings which decrease
with the distance between the spins as a power law of the type
$r_{ij}^{ - (d+\sigma )}$, with $\sigma \leq 2$. The value $\sigma=2$
corresponds to nearest-neighbor FM spin-spin interactions.

The action appropriate to describe the low-temperature properties
of the FM model (\ref{eq1}) was derived almost three decades ago \cite{dac80} and sounds as
\begin{eqnarray}
&& S\left\{ {\psi ^* ,\psi } \right\} = \sum\limits_q {\left( {r_0
+ k^\sigma   - i\omega _l } \right)} \left| {\psi \left( q
\right)} \right|^2   + \nonumber \\&&  {{\frac{T_0}{ 4V}}}
\sum_{\left\{ q_{\nu} \right\}} U\left( u_0,v_0; \left\{q _{\nu}
\right\} \right) \psi ^* ( q_1) \psi ^* ( q_2) \psi(q_3)\psi (q_4)
\end{eqnarray}
where $q\equiv (\vec k, \omega_l)$ and $U\left( { u_0,v_0;
\left\{{q_{\nu} } \right\}} \right)= \delta_{\vec k_1+\vec k_2;
\vec k_3+\vec k_4}(u_0 \delta _{\omega _{l_1 }  + \omega _{l_2 }
;\omega _{l_3 }  + \omega _{l_4 } }  $+$ v_0 \delta _{\omega _{l_1 }
;\omega _{l_3 } } \delta _{\omega _{l_2} ;\omega _{l_4 } }) $.
Here, $\psi ( {\vec k}, \omega_l )$ is a complex field related to
the in-plane magnetization, $\vec k$ denote the wave vectors (we assume a
cut-off $\Lambda =1$ related to the original spin lattice),
$\omega_l = 2 \pi l T_0 $ ($l=0,\pm 1,\pm 2,...$)
are the bosonic Matsubara frequencies and V is the volume
of the system. In Eq.~(2)  $T_0 \propto T$ ($T$ is the physical
temperature), $r_0 \propto h-[J(0)-K(0)]$, $v_0 \propto K(0)$,
where $J(0)$ and $K(0)$ are the $(\vec k =0)$-Fourier transforms
of $J_{ij}$ and $K_{ij}$ in Eq.~(1). The explicit expressions of
$T_0$ and $r_0 , u_0, v_0$ in terms of the coupling parameters
in the Hamiltonian (1),  inessential for present purposes, can be
found in Ref.~\cite{dac80}. Of course, if we put $v_0 =0$
($K_{ij} =0$ in Eq.~(1) ), $d=3$ and $\sigma =2$, the action (2)
reduces to the well-known one  for a 3-dimensional XY model in a
transverse field \cite{ger77} and for a dilute gas of hard-core
bosons  with chemical potential $\mu =-r_0$, which is believed
\cite{kawa60,kawa04,one} to be appropriate for a description of
the dimer compounds $X Cu Cl_3 $ $ (X=K, Tl)$ close to their
QCP's.

Applying the Wilson RG transformation to the action (2) and
working to one-loop approximation we obtain the equations
\begin{equation}
\begin{array}{l}
\displaystyle{dr\over{dl}} = \sigma r + {K_d \over 2}\left[ {\left( {v + 2u} \right)F_1 \left( {r,T} \right) + {vT\over{1 + r}}} \right] \\
\displaystyle{du\over dl} = \left( {\sigma  - d} \right)u - \frac{{K_d }}{2}u^2 \left[ 4{F_2 \left( {r,T} \right) + F_3 \left( {r,T} \right)} \right] - \frac{5}{2}K_d \frac{{uvT}}{{\left( {1 + r} \right)^2 }} \\
\displaystyle{dv\over dl} = \left( {\sigma  - d} \right)v - \frac{{K_d }}{2}\left( {v^2  + 4uv} \right)F_2 \left( {r,T} \right) - \frac{3}{2}K_d \frac{{v^2 T}}{{\left( {1 + r} \right)^2 }} \\
\displaystyle{dT\over {dl}} = \sigma T\left( {1 - \frac{{K_d
}}{{2\sigma }}\frac{{vT}}{{\left( {1 + r} \right)^2 }}} \right)\\
\end{array}
\end{equation}
for which the appropriate Fisher exponent is $\eta =2-\sigma$.  In
Eqs.~(3), $l$ denotes the RG rescaling parameter,
 $F_1 (r, T) = ({1/ 2})\coth[{(1+r) / T}]$, $F_2 (r,T)=-\partial F_1
(r,T)/\partial r$  and  $F_3 (r,T)={F_1(r,T)/ (1+r)}$.

The $(T_0 =0)$-critical properties can be simply obtained setting
$T=0$ in Eqs.~(3). For  case of interest $d>\sigma$, when the
quantum gaussian fixed point is stable, one finds MFA criticality
in terms of $(r_0 -r_{0c} )\propto (h-h_c)$ as $r_0 \rightarrow
r_{0c}^+$, where $r_{0c} = - ({K_d / 4d})(v_0 + 2u_0)$ localizes
the QCP. Of course, the corresponding $(T_0=0)$-critical value
$h_c$ of $h$ can be easily found from the explicit expressions of
$r_0, u_0, v_0$ in terms of the original coupling parameters.

The wilsonian classical critical regime at finite temperature for
$d<2\sigma$ can be analyzed defining $\tilde{u}=uT, \tilde{v}=vT$,
and then setting $T(l)\rightarrow \infty$ as $l \rightarrow
\infty$ in the RG equations for the new coupling parameters
$\tilde{u}$ and $\tilde{v}$.

Here, we are interested to solve Eqs.~(3) for $\sigma<d<2\sigma$
limiting ourselves to quantum or classical gaussian regime. This
can be  performed in the low-temperature limit and working to
leading order in the coupling parameters.

With $u(l)\simeq u_0
e^{-(d-\sigma)l}$, $v(l)\simeq v_0 e^{-(d-\sigma)l}$ and
$T(l)\simeq e^{\sigma l}T_0$ ($z=\sigma$ is the appropriate
dynamical critical exponent), $r(l)$ is obtained through the
non-linear relevant scaling field $g(l)=r(l)+({K_d / 4d})(v(l)
+2u(l))$ which scales as
\begin{eqnarray}
\nonumber
g\left( l \right) &=& e^{\sigma l} \left\{ {g_0  +
\frac{{K_d v_0 T_0 }}{{2\left( {d - \sigma } \right)}}
\left( {1 - e^{ - \left( {d - \sigma } \right)l} } \right) + } \right. \\
&+&\left. {  \frac{{K_d }}{2\sigma}\left( {v_0  + 2u_0 }
\right)T_0^{\frac{d}{\sigma }} \int_0^{T(l)} {dy\frac{{y^{ -
\frac{{d + \sigma }}{\sigma }} }}{{e^{\frac{1}{y}} - 1}}} }
\right\},
\end{eqnarray}
with $g_0=r_0-r_{0c}$.

We now stop the renormalization procedure at a scale $l^* \gg 1$
to be determined setting $g(l^* )\simeq 1$. Then, from Eq.~(4),
with $T_0 \ll 1$ but arbitrary $T(l^*)=e^{\sigma l^*}T_0$,
 we have  for the dimensionless inverse susceptibility
$\texttt{x} =({\chi / \chi_0})^{-1}= e^{-\sigma l^*}$, the
self-consistent equation

\begin{eqnarray}
 \texttt{x} &=& g_0 + \frac{{K_d v_0 T_0 }}{{2\left( {d - \sigma }
\right)}}\left( {1 - \texttt{x}^{\frac{{d - \sigma }}{\sigma }} }
\right)+\nonumber \\
&+& \frac{{K_d }}{{2\sigma }}\left( {v_0  +
2u_0 } \right)T_0^{d/\sigma } \int_0^{T_0 /\texttt{x}}
{dy\frac{{y^{ - \frac{{d + \sigma }}{\sigma }} }}{{e^{\frac{1}{y}}
- 1}}} .
\end{eqnarray} which contains all the relevant physical information
and allows us to explore the full quantum critical region in the
phase diagram for $d>\sigma$.

The phase boundary equation $r_{0c} (T_0)$ in the $(r_0,
T_0)$-plane, ending in the QCP, can be obtained setting
$\texttt{x}=0$ $(l^* = \infty )$ in Eq.~(5). One obtains
%
\begin{eqnarray}
g_{0c} \left( {T_0 } \right) && =r_{oc} \left( {T_0 } \right) -
r_{oc} = \nonumber \\&& - \frac{{K_d v_0 T_0 }}{{2\left( {d -
\sigma } \right)}} -
 \frac{{K_d A_{d,\sigma } }}{{2\sigma }}\left( {v_0  + 2u_0 } \right)T_0^{{d}/{\sigma }} ,
\end{eqnarray}
with $A_{d,\sigma}= \Gamma(d/\sigma )\zeta (d/\sigma )$. It is
worth noting that, if one assumes $v_0=0$, we have $r_{0c}
(T_0)-r_{0c} \propto \Delta H_c(T) = h_c (T)-h_c \propto T^{d/ \sigma} $ and hence
the result $\psi={d/ \sigma}$   for the phase boundary or shift
exponent defined as $\Delta H_c(T)\propto T^\psi$. For $d=3$ and  $\sigma=2$,
we find, as expected \cite{kawa60,kawa04}, $\psi=3/2$.
 In contrast, with $v_0 \neq 0$ and for
sufficiently low temperature, Eq.~(6) yields $\psi =1$ which is a
universal value independent of $d$ and  the interaction parameter
$\sigma$. Thus, the structure of the phase boundary  equation (6)
close to the QCP suggests the existence of a crossover temperature
$T^*_0$ between two different regimes characterized by the values $d
/ \sigma$ and 1 of the exponent $\psi$ decreasing the temperature.
By inspection of Eq.~(6), one immediately sees that (with small
$S^z-S^z$ coupling)
%
\begin{equation}
T^*_0  = \left[ {\frac{{\sigma v_0 }}{{\left( {d - \sigma }
\right)A_{d,\sigma } \left( {v_0  + 2u_0 } \right)}}}
\right]^{\frac{\sigma } {{d - \sigma }}} \sim [K(0)]^
{\frac{\sigma } {{d - \sigma }}},
\end{equation}
and the mentioned crossover is described by the effective exponent
%
\begin{equation}
\psi _{eff} \left( \tau  \right) = \frac{{\partial \ln g_{0c} \left(
\tau  \right)}}{{\partial \ln  \tau }} = \frac{{1 + \left(
{\frac{d}{\sigma }} \right)\tau ^{\frac{{d - \sigma }}{\sigma }}
}}{{1 + \tau ^{\frac{{d - \sigma }}{\sigma }} }},
\end{equation}
with $\tau = {T_0/ T^*_0}$ and $1\leq \psi_{eff} (\tau)\leq {d/
\sigma}$. For case $d=3$ and $\sigma =2$, Eq.~(8) yields $ \psi
_{eff} \left( \tau  \right) =\left [1 + ( 3/2) \tau ^{1/2}\right
]/\left (1 + \tau ^{1/2}\right)$, with $1\leq \psi_{eff} \leq 3/2$.

Although all the quantum critical properties in the gaussian
region close to the QCP can be obtained solving the
self-consistent Eq.~(5), we limit ourselves to the  experimentally
relevant case $r_0 =r_{0c}$  $(h=h_c) $ as $T_0 \rightarrow 0$. A
solution exists for $T_0 e^{\sigma l^*} =T(l^* )\gg 1$ and we find
%
\begin{equation}
\begin{array}{l}
e^{ - \sigma l^* }  \simeq \frac{{K_d v_0 }}{{2\left( {d - \sigma
} \right)}}T_0  + \frac{{K_d A_{d,\sigma } }}{{2\sigma }}
\left( {v_0  + 2u_0 } \right)T_0^{\frac{d}{\sigma }} \\
\approx \left\{ {\begin{array}{l}
   {\displaystyle\frac{{K_d v_0 }}{{2\left( {d - \sigma } \right)}}T_0\;, \qquad T_0  < T_0^* }  \\
   {\displaystyle\frac{{K_d A_{d,\sigma } }}{{2\sigma }}\left( {v_0  + 2u_0 } \right)
     T_0^{\frac{d}{\sigma }}\;,\quad T_0  > T^*_0 }.  \\
\end{array}} \right.
\end{array}
\end{equation}
This implies that also for susceptibility $\chi\sim e^{\sigma
l^*}$ and correlation length $\xi \sim \chi^{1/ \sigma}\sim e^{
l^*}$ one can define effective exponents yielding $ \gamma_{eff}
\left( \tau  \right) = \psi _{eff} \left( \tau \right)$ and
$\nu_{eff} \left( \tau  \right) = (1/\sigma )\psi _{eff} \left(
\tau \right)$. For the realistic values $d=3$ and $\sigma =2$, one has ($\chi\sim
T_0^{-1} , \xi \sim T_0^{-{1 / 2}}$) and ($\chi\sim T_0^{-{3 / 2}}
, \xi \sim T_0^{-{3 / 4}}$) for $T_0 < T^*_0$ and $T_0 > T^*_0$,
respectively.
The low-$T_0$ behaviors of other quantities
can be obtained using the standard RG machinery.

The previous results, and in particular the phase
boundary  exponent $\psi$  (or, more properly, $\psi_{eff} (\tau)$
in our RG scenario),
may constitute a good basis for comparisons
with available and  possible future experimental data.  Indeed our RG analysis
suggests that, close to the QCP, one must expect boson-like
behavior ($\psi ={d/\sigma} $) only above a characteristic
temperature $T^*\propto T^*_0 (v_0, u_0)$ depending on the coupling parameters
in the original effective FM XXZ spin model. Decreasing the physical
temperature below $T^*$, a crossover takes place to a different
regime where the phase boundary  is characterized by a different
universal  exponent $\psi =1 $. Of course, the amplitude of the crossover
region reduces to zero decreasing $v_0$. In particular, when $v_0=0$ or negligibly small,
only the exponent $\psi=d/\sigma$ should be observable. In contrast, if
the crossover region is sufficiently large, only the linear behaviour in $T$
is expected at sufficiently low temperature.

A  situation of this type seems to occur
for the doped dimer compound $Tl_{1-x}K_x Cu Cl_3$ for which
the effect of randomness on field induced magnetic ordering
has been recently investigated through low-temperature specific heat
measurements for potassium concentration in the interval
$0\leq x < 0.22$ \cite{shi04}.
The experimental predictions for the phase
boundary exponent $\psi$, relevant for our purposes, can be summarized as follows.

First, a reevaluation of  $\psi$ for $x=0$ yields
the value $\psi=1.67$ which is close to $\psi_{\rm BEC}=3/2$ derived, for compound $TlCuCl_3$,
from the theory of Bose-Einstein condensation of triplet excitations.

Next, a systematic study for $x\neq0$ shows that the disorder
produces a sensible change in the critical scenario. In particular, the phase boundary
is accurately measured for temperature below 2 K and the relevant features are: \\
i)  the exponent $\psi$ decreases systematically increasing $x$ from the
($x=0$)-value $\psi=1.67$ and tends to $\psi=1$ for $x>0.1$. This corresponds to an
enlargement of the phase boundaries obtained for different values of $x$;\\
ii) the phase boundary observed for $x>0.1$ is almost a linear function of temperature.
These findings are clearly shown in Fig.~4 of Ref.~\cite{shi04}.

To explain these results, it was conjectured in  Ref.~\cite{shi04} that a Bose glass of
triplons, in the sense argued by Fisher et al. \cite{fisher89}, could be an appropriate
model to describe properly the quantum criticality of $Tl_{1-x}K_xCuCl_3$ with
$x\neq0$. Nevertheless, the  Bose glass theory, as applied to this compound close to the
QCP, is expected to predict \cite{shi04} a phase boundary exponent
$\psi\leq 1/2$ as $T\to0$, in drastic disagreement with the available measurements.
To solve this puzzle
it was also suggested that a crossover from the convex form ($\psi>1$) to the
concave form ($\psi<1$) of the phase boundary should occur
very close to the QCP.
However, this type of crossover is not observed  in the experiment  and
hence the problem remains an open question.

The results here obtained allow us to speculate that the enlargement of the phase
boundaries and the almost linear dependence of the critical field on
temperature for $x>0.1$ \cite{shi04}
can be simply interpreted in terms of the previous RG scenario involving the
characteristic crossover temperature $T^* \sim v_0^{\sigma/(d-\sigma)}$ with
$\sigma=2$ and $d=3$. To show this, we adopt for $Tl_{1-x}K_xCuCl_3$ the idea
that the doping, as frustration, induces an effective FM coupling between
spin dimers. Furthermore we assume, reasonably, that $v_0$, and hence $T^*$,
vanishes  or its effect becomes negligible when  $x$ decreases to
zero (this lies on  the feature that the action (2), with $v_0=0$, appears to work for
the dimer compound $TlCuCl_3$ \cite{kawa60,kawa04}).

For $x=0$  our analysis predicts that the phase boundary behavior
for $T\to 0$ is characterized by the Bose exponent $\psi=3/2$,
as expected for $TlCuCl_3$ \cite{one}.
For $x\neq0$, the effective FM $S^z-S^z$ coupling in the Hamiltonian (1) becomes
active
 and a crossover is expected to occur from $\psi=3/2$
to $\psi=1$ decreasing the temperature to zero across $T^*$. When $x\lesssim 0.1$ \cite{shi04}, the
``temperature window'' $(0,T^*)$ starts to be experimentally accessible and a
$\psi \lesssim 3/2$ is measured. Increasing $x$ (above 0.1), this window becomes larger and
larger, the temperature can be decreased to zero
working below $T^*$ and one recovers the exponent $\psi=1$
which  characterizes the  linear temperature behavior of the
phase boundary observed in the experiment close to the QCP. Besides, since for $T\to0$ it is found
$3/2\geq \psi \geq 1$ increasing $x$, one can argue that an enlargement of the phase boundaries
takes place, again in agreement with the experimental data \cite{shi04}.

From our RG predictions one can obtain also some insight into the $x$-dependence
of the crossover temperature $T^*$ as given by Eq.~(7)
 in terms of the effective longitudinal
spin coupling $K(0)$, with $v_0\propto K(0)$ \cite{dac80}. Indeed,
since one expects that $v_0=0$ for $x=0$ \cite{kawa60,kawa04}, we
can assume plausibly that $v_0(x)\propto K(0)\sim x$ for small
potassium concentration. Hence, Eq.~(7) yields $T^*\sim
x^{\sigma/(d-\sigma)}$ and, for $d=3$ and $\sigma=2$, one has
$T^*(x)\sim x^2$, result which could be tested experimentally.
From a theoretical point of view, these results  should be
extracted from an appropriately chosen effective spin dimer model
and then using the explicit relations between the bare couplings
and the effective FM ones obtained by mapping \cite{tac70} the
original Hamiltonian in the FM XXZ model (1). A study of this not
easy problem is in progress and we plan to present a quantitative
scenario in a future work.

In any case, further reliable experimental and theoretical studies
on ferromagnetically coupled structural spin units are desirable
and we hope that this preliminary contribution will stimulate
 more intensive  research activity on the subject.


\begin{thebibliography}{99}
\bibitem{one} Wwhen $v_0\neq 0$, this
action  is no more  adequate to describe the quantum criticality
of an AFM XXZ model \cite{dac80}.
\end{thebibliography}
\end{document}